\documentclass[10pt,letterpaper]{article}

\usepackage{opex3}% Optics Express package
\begin{document}
\title{Non-trivial scaling of self-phase modulation and three-photon absorption in III-V photonic crystal waveguides}
\date{\today}

\author{Chad Husko$^{1,2*}$, Sylvain Combri\'{e}$^1$, Quynh Vy Tran$^1$, Fabrice Raineri$^{3,4}$, Chee Wei Wong$^2$, Alfredo De Rossi$^{1*}$}
\address{$^1$Thales Research and Technology, Route D\'epartementale 128, 91767 Palaiseau, France \\ $^2$ Optical Nanostructures Laboratory, Columbia University, New York, NY 10027 USA \\ $^3$ Laboratoire de Photonique et de Nanostructures (CNRS UPR 20), Route de Nozay, 91460 Marcoussis, France \\$^4$ Universit\'e D. Diderot, 75205 Paris, France \\ 
\url{cah2116@columbia.edu, alfredo.derossi@thalesgroup.com}}

\begin{abstract}
\noindent
We investigate the nonlinear response of photonic crystal waveguides
with suppressed two-photon absorption. A moderate decrease of the
group velocity ($\sim$ c/6 to c/15, a factor of 2.5) results in a dramatic
($\times$ 30) enhancement of three-photon absorption well beyond the expected scaling, $\propto{1/v_g^3}$. This non-trivial scaling of the effective nonlinear coefficients results from pulse compression, which further enhances the optical field beyond that of purely slow-group velocity interactions. These
observations are enabled in mm-long slow-light photonic crystal
waveguides owing to the strong anomalous group-velocity dispersion and positive chirp. Our numerical physical model matches measurements remarkably.
\end{abstract}

\ocis{(130.5296) Photonic crystal waveguides, (190.4180) Multiphoton processes, (190.3270) Kerr effect, (190.4400) Nonlinear optics, materials, (250.4390) Nonlinear optics, integrated optics} % 

%\bibliography{OptExp_slowLightSPM}		% Produces the bibliography via BibTeX.`

\section{Introduction}
\indent Slow-light nonlinearities have been remarkably observed through quantum coherence and interference in atomic systems, with group velocities of tens of meters per second and sub-100 kHz bandwidths\cite{harris1997, hau1999, longdell2005, deng2003}. In solid-state systems such as photonic crystal cavities \cite{yang2009} or waveguides \cite{baba_nature2008, krauss2008, vlasov2005}, modest slow group velocities of down to c/300 in comparison has been achieved but possess THz bandwidths for chip-scale optical signal processing \cite{foster2008_nature, pelusi2009}. The strong ab-initio structural dispersion in slow-light photonic crystals waveguides (PhCWGs) not only gives rise to localized modes\cite{topolancik2007,engelen2008,baron2009}, but also dramatic enhancement of resonant and non-resonant nonlinearities\cite{soljacic2004,mcmillan2008}. Third-order processes have also recently been observed for slow-light \cite{okawachi2005}, self-phase modulation (SPM) \cite{siviloglou2006, monat2009, inoue2009}, and third-harmonic generation in optical microstructures \cite{carmon2007, markowicz2004, corcoran2009}. Here we present the first observations of self-phase modulation limited only by three-photon-absorption as well as evidence for pulse compression in slow-light GaInP photonic crystal waveguides. In contrast to previous self-phase modulation work in PhCWG\cite{baron2009,monat2009, inoue2009}, which was limited by the third-order two-photon absorption (TPA), here the nonlinear loss term is the fifth-order three-photon absorption. The different orders of the desirable and undesirable nonlinearities, along with tight modal confinement of the photonic crystal, lead to the observation of novel optical effects. Remarkably the three-photon absorption process demonstrates a 30-fold enhancement, and departs from the expected ($1/v^3_g$) scaling, even when taking into account slow-light disorder scattering. Pulse compression was further observed in our positive chirp due to the interaction of the Kerr effect with anomalous group velocity dispersion(GVD). These series of measurements show excellent match with our numerical simulations including group-velocity-dependent nonlinearities and losses, and measured group velocity dispersion through optical low-coherence reflectometry~\cite{combrie07}. 

\indent The origin of slow-light in photonic crystals arises from coherent Bragg reflections due to the in-plane periodic PhCWG lattice, leading to an ultra-flat dispersion of the transverse electric (TE) field. At these so-called slow-light frequencies, the light effectively travels slowly through the lattice via multiple Bragg reflections, leading to an enhanced local field density. The local field enhancement scales inversely with the group velocity, thus decreasing the threshold of intensity-dependent nonlinear effects such as Kerr, multi-photon absorption, or Raman scattering \cite{baba_nature2008,soljacic2004,mcmillan2008}. The combination of the engineered group velocity enhancement, along with the small modal effective area of PhCWGs ($\approx 2 \times 10^{-13}$ m$^2$), yields outstanding control over the optical modes.

\indent The optical Kerr-effect has long been studied for its utility in nonlinear all-optical devices~\cite{soljacic2004, stegeman1993}. In addition to the desirable Kerr term, each material also has a nonlinear multi-photon absorption term. This fundamental material property induces several detrimental effects, such as: (a) limiting the Kerr-induced phase-shift, also called self-phase modulation (SPM); (b) inducing nonlinear and free-carrier absorption losses; (c) distorting the pulse via free-carrier dispersion\citation{yin2007}; as well as (d) restricting the spectral range of any potential nonlinear optical devices such as all-optical switches\cite{kang1994}. In fact, several early studies point out that the wavelength range of the two-photon absorption (TPA) tail depends strongly on the quality of the molecular beam epitaxy (MBE) sample growth~\cite{kang1994,villeneuve1993,aitchison1997}. These fundamental obstacles must be overcome in order to achieve practical nonlinear devices, such as all-optical switches, on-chip. With this insight, we selected the material system GaInP ($E_g$=1.9 eV) for its desirable nonlinear properties. With an energy bandgap at least 300 meV above the TPA range ($E_g >2\hbar\omega$), the GaInP sample investigated here is well outside the range of potential defect states and the sole nonlinear loss mechanism is three-photon absorption (ThPA). Importantly, ThPA is negligible at the optical intensities required for the Kerr-effects ($>\pi$ phase-shifts) observed experimentally in this paper. The complete suppression of TPA and small impact of ThPA on the GaInP PhCWG open the possibility of a significant spectral window for all-optical signal processing\cite{kang1994}.
%%%%%%%%%%%%%%%%%%%%%FIGURE 1%%%%%%%%%%%%%%%%%%
\begin{figure}[h]
\centering
\includegraphics[width=10cm]{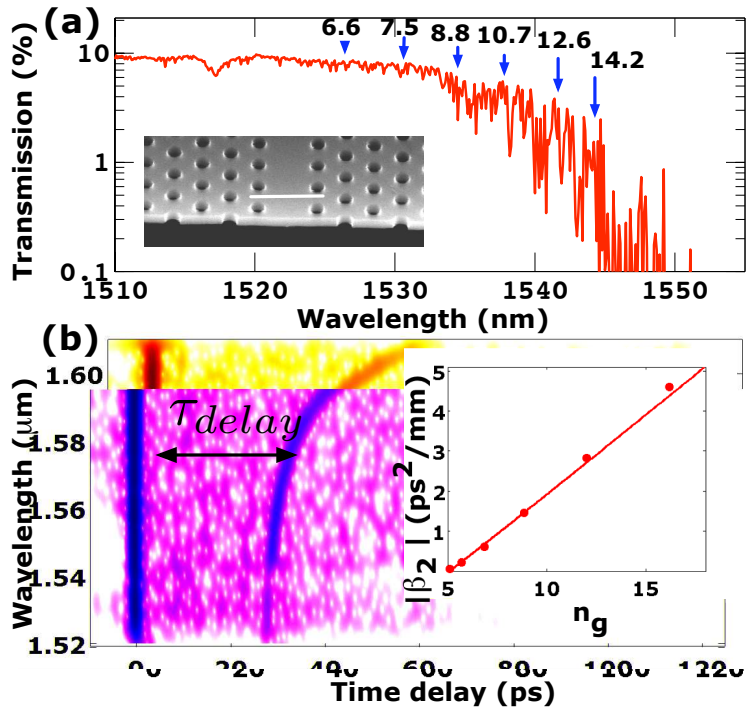}
\caption{(Color online) Linear properties: (a) Measured PhCWG transmission and corresponding $n_g$ of 6.6, 7.5, 8.8, 10.7, 12.6, 14.2, respectively, from low-coherence reflectometry after Refs. \cite{combrie07,parini2008}. Inset: SEM image with scale bar of 1 $\mu$m. (b) Sample optical low-coherence reflectometry (OLCR) data used to extract $n_g=c\frac{\tau_{delay}}{L}$. Inset: Extracted group indices versus derived GVD coefficients.}
\label{fig:figure1}
\end{figure}
%%%%%%%%%%%%%%%%%%%%%%%%%%%%%%%%%%%%%%%%%%%%
\section{Experiment}
\subsection{Sample details and linear characterization}
\indent Our PhCWG sample is a W1 GaInP membrane of 190 nm thickness, with a hexagonal lattice constant $a$ of 480 nm, hole radius $r\approx0.23a$, and 1.5 mm length. The fabrication has been described elsewhere~\cite{combrie_05} and the linear PhCWG transmission is illustrated in Figure \ref{fig:figure1}(a). We carefully designed integrated mode-adapters~\cite{Tran2009} to reduce losses to $\sim$ 5 dB/facet at 1526 nm~\cite{weidner07} and suppress Fabry-Perot oscillations, though disordered propagation due to the lattice is clearly seen at the slow-light onset region~\cite{parini2008}. Optical low-coherence reflectometry~\cite{combrie07,parini2008} [Fig. \ref{fig:figure1}(b)] was used to extract the group indices. and the group velocity dispersion (GVD) [Fig. \ref{fig:figure1}(b) inset] computed using complete 3D planewave expansion~\cite{mpbCode}, with measurement-consistent group indices.
\subsection{Nonlinear characterization - three-photon absorption}
\indent For the nonlinear measurements, we employed a wavelength-tunable mode-locked fiber laser (PriTel) with $\approx$ 3 to 5 ps pulses~\cite{fiberLaserNote} (characterized with autocorrelator) at 22 MHz with transverse-electric (electric-field in-plane) polarization. The output pulse monitored with a spectrum analyzer and an oscilloscope. The peak power coupled to the PhCWG $P_{c}$ is defined as: $P_{c}=\eta K_cP_{in}$, where the input peak power $P_{in}$ (output power $P_{out}$) includes the objective loss (2 dB), mode mismatch (2 dB) $\eta$~\cite{Tran2009}, with the coupling coefficient, $K_c$. The insertion loss due to disorder varies with slow-light ~\cite{kuramochi_PRB2005}; the values of which are extracted directly from Eqn. \ref{ThPA_extraction} below.
%%%%%%%%%%%%%%%%%%%%%FIGURE 2 %%%%%%%%%%%%%%%%%%
\begin{figure}[h*]
\centering
\includegraphics[width=10cm]{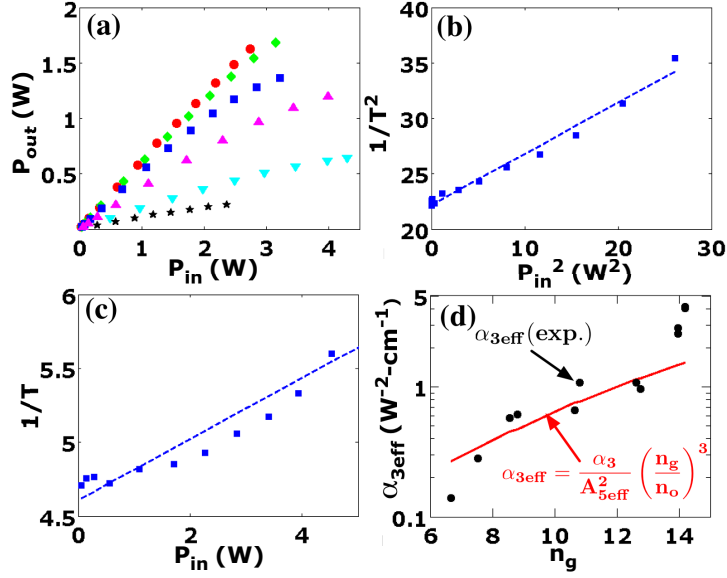}
\caption{(Color online) Nonlinear measurements: (a): $P_{in}-P_{out}$ depending on the group velocity. The increased curvature corresponds to larger nonlinear absorption due to slow-light enhanced three-photon absorption (ThPA) at longer wavelengths. Key: 1526 nm(circles), 1530nm(diamonds), 1534 nm(squares), 1538 nm(upward triangles), 1541 nm(downward triangles), and 1544 nm(stars).  (b) Sample plot of the inverse transmission squared ($1/T^2$) versus $P_{in}^2$ at 1534 nm ($n_g$=8.8). The points are experimental data and the line is the best fit to extract the effective ThPA coefficient $\alpha_{3eff}$. The key is the same as in (a). (c) Example inverse transmission ($1/T$) versus $P_{c}$ plot depicting the mismatch of (and negligible) two-photon absorption in our slow-light GaInP PhCWGs, for the same experimental data of panel (b). (d) Extracted $\alpha_{3eff}$ (black dots) versus group index with the expected scaling of ThPA (solid red curve)}
\label{fig:fig2} 
\end{figure}
%%%%%%%%%%%%%%%%%%%%%%%%%%%%%%%%%%%%%%%%%%%%

\indent In Fig. \ref{fig:fig2}(a), we first examined $P_{in}-P_{out}$ at different group velocities and same input power range to illustrate the enhanced nonlinear absorption processes. At the largest group velocities, $P_{in}-P_{out}$ is linear while at smaller group velocities the output power begins to saturate, indicating the distinct onset and slow-light enhancement of ThPA, an intensity dependent loss mechanism, in the slow-light regime. The nonlinear propagation equation ($\partial P /\partial z = -\alpha P - \alpha_{3} P^3$) with symmetric input-output coupling gives:
\begin{equation}
\label{ThPA_extraction}
\frac{1}{T^2}=\frac{P_{in}^2}{P_{out}^2}=\frac{1}{\eta^2K_c^4 e^{-2\alpha L}}+\frac{2\alpha_{3eff} L_{3eff} }{K_c^2 e^{-2\alpha L}}  P_{in}^2,
\end{equation}
where $\alpha_{3eff}=\frac{\alpha_3}{A_{5eff}^2}\left(\frac{n_g}{n_o}\right)^3$ includes the 5th-order modal effective area, and $L_{3eff}=(1-e^{-2 \alpha L})/(2\alpha)$. The bulk $\alpha_{3}$ coefficient, $\sim 6 \times 10^{-26} m^3/W^2$, was calculated from a well-cited model\cite{wherrett1984}, experimentally verified for the similar AlGaAs material in Ref. \cite{aitchison1997}.

\indent
In Fig. \ref{fig:fig2}(b), we illustrate an example plot of $1/T^2$ versus $P_{in}^2$ to extract the effective nonlinear ThPA coefficient and coupling factor $K_c$ at a particular group index $n_g$ of 8.8. We note the material-dispersion of the nonlinear susceptibilities is negligible within our measurement range. An example two-photon absorption ($1/T$ versus $P_{in}$) analysis~\cite{aitchison1997} [Fig.\ref{fig:fig2}(c)] shows clear mismatch of the experimental data (blue squares) with the TPA fit (dashed line). Similar fits occur for all group velocities investigated, thus confirming the sample experiences only three-photon absorption. We note that, for the group velocities we examined, the linear loss coefficient $\alpha$ has been observed to scale \cite{engelen2008, kuramochi_PRB2005,hughes2005} approximately as $n_g^2$, with our measured values of 1 dB/mm at 1526 nm~\cite{combrie_opex06}. In the semilog plot of Fig.~\ref{fig:fig2}(d), we illustrate the extracted three-photon $\alpha_{3eff}$ values in the slow-light regime. In this first observation of three-photon absorption in PhCWGs, we further emphasize that a surprising \emph{30-fold enhancement} in three-photon coefficient from 0.15 to 4 $W^{-2}cm^{-1}$ was observed  with $n_g$ increasing only from 6.6 to 14.2. While the field enhancement of nonlinear absorption due to slow-light can be predicted from the material($\alpha_3$) and device parameters ($n_g$) and is known to obey a prescribed trend\cite{soljacic2004}, here the deviation in the \textit{shape} of the trajectory with group index ($n_g$) is markedly different than the scaling predicted from exclusively slow-light effects, an indication that some other field enhancement effect is involved. Moreover, this enhancement is by far faster than the predicted scaling from the definition of the ThPA coefficient \cite{AeffNote2}, $\frac{\alpha_3}{A_{5eff}^2}(\frac{n_g}{n_o})^3\approx5.8$ enhancement expected from simple slow-light effects. We now further elaborate on this point in terms of pulse self-phase modulation due to the optical Kerr effect.\\

\subsection{Nonlinear characterization - Kerr-effect}
\begin{figure}[h]
\centering
\includegraphics[width=13.5cm]{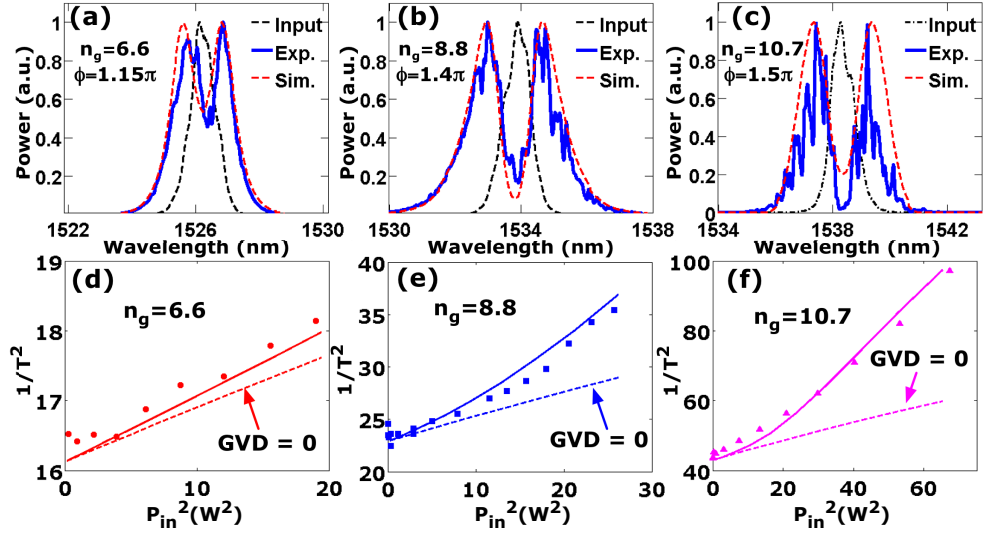}
\caption{ (Color online) Comparison of experimental and theoretical spectra for different wavelengths of 1526 nm ($n_g$=6.6) (a), 1534 nm ($n_g$=8.8) (b), and 1538 nm ($n_g$=10.7) (c), with chirped $sech^2$ input. Experimentally derived parameters, also used in simulations, are shown in Table \ref{tab:parameters}. (d) Experimentally $1/T^2$ versus $P_c^2$ for 1526 nm($n_g$=6.6). The solid line shows the simulated results including GVD, while the dashed line are simulation results without GVD. (e),(f) Same as (c) for but with (e) 1534 nm($n_g$=8.8) and (f) 1538 nm($n_g$=10.7). The strong upward bend of the curve indicates enhanced nonlinear absorption beyond conventional slow-light scaling, triggered from pulse compression.}
\label{fig:expSpectra} 
\end{figure}

\indent To further illuminate this nontrivial scaling, we measured the output spectra at different slow group velocities [1526, 1534, and 1538 nm in Figs. \ref{fig:expSpectra}(a), (b), and (c) respectively]. At increased input intensities, the pulses undergo stronger Kerr self-phase modulation with the associated spectral broadening. In each plot, we show examples of low (dot-dashed black line) and high ($>$2 W; solid blue line) power spectra. For group indices less than $\sim$ 10, a clear increase in spectral width was observed with increasing group index, while maintaining the same coupled peak power. The fine-peaked structure in the pulse spectra is due to disorder-enhanced scattering~\cite{parini2008}. The high-power spectra exhibit symmetric double-peak structures characteristic of Kerr-induced SPM, where the pulse spectral symmetry rules out free-carrier dispersion generated by ThPA~\cite{monat2009} or third-order dispersion~\cite{dadap2008, ding2008}. 
\begin{table}
\centering
\caption{\label{tab:parameters} Parameters used in numerical simulations}
\begin{tabular}{lllll}
Parameter &Value&Value&Value&Value.\\ 
\hline
Wavelength     $\lambda (nm)$   &  $1526$  &  $1530$   &  $1534$ &  $1538$\\
Pulse duration (FWHM)     $\tau (ps)$   &  $5.0$  &  $4.4$ &  $4.5$ &  $3.0$ \\
Chirp     $C$   &  $1.3$ &  $1.1$ &  $1.2$ &  $0.2$     \\
Time-bandwidth product     $\Delta\nu\,\tau$   &  $0.74$ &  $0.69$ &  $0.69$ &  $0.33 $     \\
Peak Power     $P_c (W)$   &  $1.85$ &  $2.2$ &  $2.25$ &  $2.35$  \\
Group index     $n_g$   &  $6.6$ &  $7.5$ &  $8.8$ &  $10.7$\\
GVD     $\beta_2 (ps^2/mm)$   &  $-0.6$ &  $-0.97$  &  $-1.4$ &  $-2.2 $   \\
Eff. ThPA  $\alpha_{3,eff} (cm^{-1}W^{-2})$   &  $0.24$  &  $ 0.40$ &  $0.61$ &  $0.66 $\\
SPM gain    $\gamma (cm^{-1}W^{-1})$ &  $6.75 $ &  $7.96 $ &  $9.7$ &  $13.0 $ \\
\end{tabular}
\end{table}
 
\indent To further quantify our experimental results, we performed numerical simulations of the nonlinear Schr\"odinger equation\cite{agrawalNLoptics}:
\begin{equation}
\label{NLSE}
\frac{\partial A}{\partial z} = -\frac{\alpha}{2}A-\frac{\alpha_{3eff}}{2}|A|^4A +ik_o\frac{n_{2}}{A_{3eff}}|A|^2A -i\frac{\beta_2}{2} \frac{\partial^2A}{\partial t^2}
\end{equation}
where $A$ is the pulse envelope amplitude with $P_{c}=|A|^2$, $\beta_2$ [ps$^2$/mm] the GVD parameter, $z$ is the propagation direction, and $t$ defined in a moving reference frame. The effective nonlinear coefficients are defined according to Ref. \cite{Bhat_PRE2001, AeffNote} and modes from planewave expansion~\cite{mpbCode}. We computed $A_{5eff}$ for the fifth-order nonlinearity, ThPA,  and $A_{3eff}$ for the third-order Kerr nonlinearity. Our calculations show that $A_{5eff}\approx0.75A_{3eff}$ over the wavelength range and waveguide geometry examined here. The third-order area $A_{3eff}$ ranged from $2$ and $2.6\times 10^{-13}~m^2$ for the wavelengths examined, an increase far smaller than the slow-light scaling. Assuming a conservative absorbed energy per pulse due to ThPA of 20\%, for a ($\sim$4 ps) this corresponds to $\sim$pJ or less distributed along the full length of the waveguide (1.5 mm). At the repetition rate (22 MHz) of the laser, the average absorbed power would then be on the order of tens of $\mu$W. Thus at the power levels used in the experiment, free-carrier effects contribute negligibly to the pulse dynamics in the GaInP material. Positive chirped (chirp parameter $C$) hyperbolic-secant input pulses\cite{agrawalNLoptics}, such as those generated by the mode-locked fiber laser in the experiment, are used. The variation of the chirp over the wavelength range is a characteristic property of the laser. A Kerr $n_{2}$ coefficient of $8\times10^{-14} cm^2/W~$\cite{sheikbahae1990}, scaled with $n_g^2$, along with the experimentally-extracted values of $\alpha_{3eff}$ are used in our model. Table \ref{tab:parameters} summarizes the measured parameters in our model, with the peak power within $10\%$ of measured values (uncertainty from chirp and pulsewidth).
%%%%%%%%%%%%%%%%%%%%%FIGURE 3%%%%%%%%%%%%%%%%%%
%%%%%%%%%%%%%%%%%%%%%%%%%%%%%%%%%%%%%%%%%%%

\indent The resulting simulated spectra with chirp and GVD at three different group indices, $n_g$=6.6, 8.8, and 10.7 in Figs. \ref{fig:expSpectra}(a), (b), and (c), respectively, show remarkable matches with our measurements. To examine the effects of chirp and GVD on ThPA, in Fig.~\ref{fig:expSpectra}(d), (e), and (f) we demonstrate 1/T$^2$ versus $P_c^2$ generated strictly from experimental parameters. The complete physical model not only describes the higher-order nonlinear absorption (1/T$^2$ versus $P_c^2$), but also captures its scaling to larger values in both slope and curvature at lower group velocities. The inclusion of anomalous GVD in the model, and thus the possibility of pulse compression, not only reproduces the slight upward bending, but also rigorously and correctly predicts the increase of the effective absorption \emph{beyond} the $n_g^3$ scaling. We note that higher order dispersion is still negligible. The slight upward bending in $1/T^2$ vs. $P_{in}^2$ plot is the signature of pulse-compression. We emphasize that despite this small bending, the fit for $\alpha_{3eff}$ are quite good for all values examined. We also examined the case where GVD is set to zero as the dotted line in Figs. \ref{fig:expSpectra}(d-f). The strong deviation between the simulation when GVD is zero and experimental data clearly demonstrate that GVD is interacting strongly with the nonlinear effects. We also simulated the absence slow-light effects, that is $n_g=3.12$. In all cases, the curves were well below the experimental data, and nearly overlapped the zero GVD simulation data. These are not included here for figure clarity.

\begin{figure}[h]
\centering
\includegraphics[width=13.5cm]{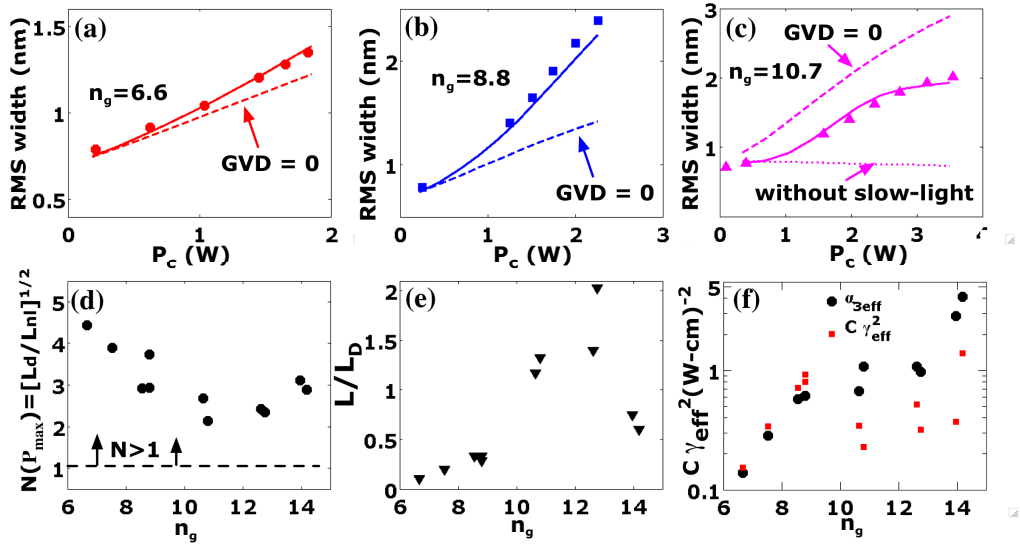}
\caption{(Color online)(a) and (b) RMS pulse broadening (nm) as a function of coupled input power $P_{c}$ at: (a) 1526 nm ($n_g$=6.6), (b) 1534 nm ($n_g$=8.8) and (c) 1538 nm ($n_g$=10.7). The points are experimental data, the solid line is simulation with GVD, and the dashed line is without GVD. We also show the case without slow-light (e.g. $n_g=3.12$) in (c) as the dashed line. (d) $N=\sqrt{L_D/L_{NL}}$ versus $n_g$ at $P_{in}-max$(W), the max peak power input into the PhCWG. For values of $N>1$, the pulse has the possibility of being compressed. (e) Plot of $L/L_D$ vs. $n_g$. In addition to $N>1$, the pulse must also propagate a minimum length, related to the dispersion length $L_D$, before compression can occur. (f) Effective nonlinear absorption, $\alpha_{3eff}$ and square of the effective SPM coefficient,$\gamma^2_{eff}$, rescaled with a suitable constant, $C$. The local field enhancement of the two effects scales as predicted. The experimental values demonstrate non-trivial scaling.}
\label{fig:spmData}  
\end{figure}

\indent To further examine the impact of pulse compression on the slow-light Kerr nonlinearity, we quantified the measured SPM-induced spectra broadening with: $\Delta\lambda^2=\frac{\int{(\lambda-\lambda_0)^2\,S\,d\lambda}}{\int{S\,d\lambda}}$, where $S$ is the lineshape. Illustrative measurements of spectral broadening at $n_g$=6.6, 8.8, and 10.7 for increasing coupled power are shown in Fig. \ref{fig:spmData}(a), (b), and (c), respectively. While SPM alone (dashed line) produces a linear slope that has a slight downward curvature due to higher-order absorption for large $n_g$, the spectral widths including the effects of GVD mark a significant departure from a linear slope, as observed both experimentally (markers) and numerically(solid line). Fig. \ref{fig:spmData}(c) in particular demonstrates a saturation of the spectral broadening at larger input powers. This is confirmed to arise from a combination of ThPA and GVD in the numerical model. We additionally plot the case without slow-light effects as the dashed line. Clearly slow-light has a strong impact on the spectral broadening, as the behavior is completely linear. Similar results occur at the other wavelengths.\\
%%%%%%%%%%%%%%%%%%%%%FIGURE 4%%%%%%%%%%%%%%%%%%
%%%%%%%%%%%%%%%%%%%%%%%%%%%%%%%%%%%%%%%%%%%
\subsection{Discussion of non-trivial scaling}
\indent Thus far we have observed trends in both our ThPA and Kerr measurements that do not obey conventional slow-light scalings. Clearly a field enhancement effect beyond slow-light enhancement is occurring in the sample. Based on the necessity of including GVD in our modeling, and with knowledge that pulse compression (resulting from the interaction of the Kerr self-phase modulation and anomalous dispersion) causes increased peak power of an optical pulse~\cite{agrawalNLoptics}, we investigated pulse compression as a source of the additional local field in our PhCWG. We first computed the dispersion length, $L_D$, ($\sim$ 14 mm at 1526 nm and 1.3 mm at 1538 nm) and the nonlinear length, $L_{NL}$, ($\sim$ 1.3 mm at 1526 nm and 0.45 mm at 1538 nm)~\cite{agrawalNLoptics} at the various pulse center frequencies of our slow-light experiments. The results, $N=\sqrt{L_D/L_{NL}}$, are plotted in Fig. \ref{fig:spmData}(d). For $N>1$ and $\beta_{2}C < 0$, pulse compression is possible\cite{agrawalNLoptics}. There is, however, an additional criterium to observe pulse compression. The requirement is that the pulse must propagate a minimum distance compared to $L_D$, the dispersion length, and depends on $N$.  This implies that the physical device length must be comparable to $L_D$, typically at least $L_D/2~$\cite{agrawalNLoptics}. Figure \ref{fig:spmData}(e) demonstrates the second key parameter, the $L /L_D$ ratio, for the experimental data. While $N$ is always greater than one, the dispersion length here decreases such that $L /L_D$ increases from $L/L_D =0.1$ to $L_D=1.4$ as $n_g$ goes from 6.6 to 12.6. Thus while the pulses of the lowest three group indices $n_g$ experience some compression ($N>1$ and $L/L_D \approx 0.1-0.3$), the larger $n_g$ values experience more compression since they possess both $N>1$ and ratios of $L/L_D > 0.5$. Thus both criteria have been met. This accounts for the faster than predicted scaling of $\alpha_{3eff}$.\\

\indent We now utilize the extracted broadening data to determine the effective SPM coefficient $\gamma_{eff}$ from the nonlinear phase $\phi_{max} = \gamma_{eff} P_c L_{eff}$ and spectral broadening: $\frac{\Delta\lambda}{\Delta\lambda_0} = \sqrt{1+\frac{4}{3\sqrt{3}}\phi_{max}^2}$. Here the effective length $L_{eff}$ is limited only by ThPA and hence must be defined differently from the typical linear loss parameter~\cite{effLengthNote}. The enhancement of the three-photon absorption and SPM broadening are related through the field enhancement factor, $\kappa$, relating the propagating pulse power to the electric field energy density: $|E|^2 = \kappa P$. We thus have $\gamma_{eff} \propto \kappa$ and $\alpha_{3,eff} \propto \kappa^2$, implying that $\alpha_{3eff}$ scales as $\gamma_{eff}^2$. In Fig. \ref{fig:spmData}(f), we examine $\alpha_{3eff}$ (black circles) and $\gamma_{eff}^2$ (red squares; scaled by a constant $C$) -- the curves are remarkably similar, highlighting unambiguously that the non-trivial scalings of Kerr and ThPA have a common physical origin, that of strong field enhancement consistent with pulse compression.
\section{Conclusion}
\indent We have demonstrated nonlinear scaling of the nonlinear enhancement beyond the limits of slow-light photonic crystal waveguides. This non-trivial scaling has been analyzed with experimental data of both self-phase modulation and three-photon absorption and further reinforced with a nonlinear propagation model. The origin of this non-trivial scaling beyond the slow-light regime is derived from an additional local field enhancement due to pulse compression. As Kerr is a third-order effect in field, while ThPA is a weaker fifth-order nonlinear effect, dispersion engineering to include pulse compression can greatly reduce the threshold of SPM, while carefully managing ThPA and its undesirable properties. The ability to engineer PhCWG dispersion\cite{vlasov2005}, such as with low group velocity dispersion~\cite{kubo2007} to precisely control the pulsewidth, while completely suppressing TPA, make GaInP PhCWGs promising for compact broadband ultra-fast optics.\\
\\
\textbf{Acknowledgments} \\
\\
\noindent This work was supported by the European Commission (project GOSPEL), the Fulbright Grant (C. Husko), the National Science Foundation, and the New York State Foundation for Science, Technology and Innovation. We thank R. Gabet and Y. Jaou\"en (Telecom ParisTech) for OLCR setup and acknowledge discussions with J. F. McMillan.


\begin{thebibliography}{10}
\bibitem{fiberLaserNote}
Rigorous intensity autocorrelation measurements of the pulse were carried out
  at each wavelength and current setting.
\bibitem{AeffNote2}
A group-index dependent effective area for the slow-light PhCWG is included in
  the fifth-order nonlinear scaling, with $A_{5eff}$ defined as:
  $A_{5eff}=\frac{V_{5eff}}{a}=\bigg[\frac{(\int_{n2}{\epsilon
  |E|^2}dV)^3}{\int_{n_2}\epsilon|E|^6dV}\bigg]^{1/2}$.
\bibitem{AeffNote}
The vector definition is applicable to materials with large indices of
  refraction and tight confinement such as PhCWGs:
  $A_{3eff}=\frac{V_{3eff}}{a}=\frac{(\int_{n_2}{\epsilon|E|^2}dV)^2}{\int_{n2%
}{\frac{\epsilon}{3} (|E\cdot E|^2+2|E|^4}dV}$. The scalar definition is the
  typical: $A_{eff}=\frac{V_{eff}}{a}=\frac{(\int_{n2}{\epsilon
  |E|^2}dV)^2}{\int_{n_2}\epsilon|E|^4dV}$.
\bibitem{effLengthNote}
The often cited linear loss definition, $L_{eff}$(linear) $=(1-e^{-\alpha
  L})/\alpha$, is not applicable in our current experiment to estimate
  $\phi_{max}$. We derive a more appropriate definition for when three-photon absorption, as opposed to linear loss, is the dominant loss term: $L_{effNL}(ThPA) = \frac{1}{I_o\sqrt{\alpha\alpha_3}}
  \arctan\bigg[\frac{2L_{3eff}I_o \sqrt{\alpha\alpha_{3}}}{{1+e^{-\alpha
  L}\sqrt{1+2\alpha_{3} I_o^2L_{3eff}}}}\bigg]$, to be detailed elsewhere.
\end{thebibliography}
\end{document}